\documentclass[12pt,a4paper]{article}
\begin{document}
                                  Non commutative quantum spacetime with  topological vortex states,

                                                     and dark matter in the Universe

                                                             Ajay  Patwardhan

                                                  Physics Department, St. Xavier's college,

                                                   Mahapalika Marg,  Mumbai, 400001, India

                                                                 Abstract

Non commutative geometry is creating new possibilities for physics. Quantum spacetime geometry and post inflationary models of the universe with matter creation have
an enormous range of scales of time, distances and energy in between. There is a variety of physics possible till the nucleosynthesis epoch is reached. The use of topology and non commutative geometry in cosmology is a recent approach. This paper considers the possibility of topological solutions of a vortex kind given by non commutative structures. These are interpreted as dark matter, with the grand unified Yang-Mills field  theory energy  scale used to describe its properties. The relation of the model with other existing theories is discussed.

Introduction

The role of Geometry and Topology in understanding the physics of the universe of space time and matter has a century of work that is supported by observations. Inflationary cosmological models developed for creation of matter have included the small anisotropic distributions as given by COBE and WMAP observations. The models modified to create the voids and filamentary large scale structure of the early universe , with the ratios of dark energy, dark matter and normal matter are topics of active research.

The non linear evolution equations of the early universe  give  rise to anisotropy and inhomogenity persisting at all length scales , these  are  observable with even  finer resolution that is  expected in subsequent experiments. These  indicate analogies with condensed matter. Topological defects and vortex like states in rapid expansion and cooling occur in condensed matter phases.

This paper develops the role of noncommutative geometry in the formation of the vortex like topological states. The typical model of the universe, a few seconds old and a few kilometers in size, is expected to have some $ 10^{80} $ particles of normal matter, and to form a smooth geometry given by Einstein's theory of general relativity ; with  $ 10^{-5} $ parts fluctuation in matter distribution as inferred from the temperature and thermal radiation microwave measurements. [References].

The models of the early universe

In the inflation models and their variations that give the later stage expanding universe models with the  correct observed properties, there are topologically non trivial states possible. A rubber sheet or expanding balloon picture with locally ' pinched and popped ' regions is suggested. These topological 'defects'could be candidates for dark matter with a long (billion years) half life for decay into normal matter. A topological defect would give rise to a graviton and a photon as it `pops out of the pinch', and gives a smooth manifold in its neighbourhood. Typical masses of $ 10^{-15} $ kilograms have been suggested, which would give with  ($ E= m c^2 $) ,gamma rays of $10^{21} $ electron volts ; generating normal matter and the graviton emitted would propagate as a gravitational wave contributing to the smoothing out of the `wrinkles' in local space time. From a $ 1:100 $ to a $1:30 $ ratio of normal to dark matter are reported in the early universe and a ratio $ 1:8 $ in galaxies and their halos. The slow decay could be a perturbative or non perturbative effect; but it involves a change in topology. Ultra heavy dark matter at the $ 10^{15} $ Gev GUT scale and its decay into gamma rays and baryon ,lepton production are known as an alternative to the baryogenesis stage of the universe.

The decaying topological defects give rise to the $ 10^{21} $ e.v. gamma ray emission  consistent with cosmic rays of highest energy gamma rays observed, and the  nearly isotropic distribution of the gamma ray bursts in the early  Universe. The dark matter in galaxy halos is the remnant of the earlier rapidly decaying stage ,possibly an exponential decrement of the number of the topological defects. The grand unification (GUT) scale of $ 10 ^{15} $ Gev , is also comparable and indicates that a single Holonomy gauge field , with gauge bosons treated as nearly massless, and represented by the single coupling constant  would describe the loop integral around the ' defect' as it decays. In theories of Quantum chromodynamics of the quark gluon plasma phase transitions and  the  inflation cosmological models with a scalar field , topological defects are possible. At a scale between GUT ( $ 10^{15} $ Gev) and of  Quantum Riemann geometry ( $10^{19} $ Gev), a non commutative structure can arise in the non linear regime of the evolution of the combined model of all the dynamics of the universe. The so called vaccuum manifold $ M -> G/H $ as a symmetry breaking with group G and subgroup H gives rise to topologically non trivial solutions.

The non commutative $ \theta $ structure with a length scale of $ 10^{-27} $ meters ; $ h/mc  $ with $ m= 10^{-15} $ kg, for the vortex like topological defect , goes over to a commutative differential geometry in the large scale. It is possible to retain the symmetric metric and connection and absence of torsion in the large scale. In a geometric interpretation of charge the continuous electromagnetic field F  has a topological index as an integral of F*F , as also for gravity of geons in geometrodynamics. The topological defects of dark matter arise in the background geometry due to the non commutative structure with a $ 10^{-27} $ length scale.  The number and  number density, mass and mass density of the dark matter as topological states in the universe are still not known. As these 'topological defect' regions are excluded from the else where smooth geometry, they do not interact with electromagnetic radiation and are 'dark' matter; better called 'dark' space time matter.

At the  Planck scale, theories of Quantum Gravity such as loop space quantum gravity and quantum Riemann geometry give a spin network representation with area and volume operators and holonomy for the fields. Quantum space time with a non commutative geometry is proposed as an alternative to theories with  other forms of exotic matter , primordial black holes and  sources of gamma bursts. The presence of vorticity can be described by a general Stokes theorem and the second exterior derivative . In the condition for exact and closed forms, the topology of the manifold plays a role; the genus number , betti numbers, Euler numbers, Hopf and Pontryagin index and other topological indexes. The non inertial vaccuum in the accelerated expanding geometry gives rise to particle creation by Bogoliubov transforms. Topologically non trivial solutions are known for a variety of non linear evolution equations. The quartic potential scalar field with a non zero vaccuum state in the standard inflation models is the starting point for various theories of early cosmology. The field could be a residue from the transition of non commutative to commutative geometry and thus have a geometrised rather than independent existence.

The Model including non commutative geometry

Consider the vector field $\chi$ that has the non commutative structure $\chi_{i} \chi_{j}  - \chi_{j} \chi_{i}  = i \theta_{ij} l^{2} $.

The spacetime coordinates are obtained as $ \chi_{i} ->  x_{i} $ in the limit of commutative geometry.

 The congugate momenta have the relation $ p_{i} p_{j}  - p_{j} p_{i} = \theta_{ij} ^{-1} l^{-2}h^{2}/4\pi^{2}  $.

 Therefore the usual quantisation rule of $ \chi_{i} p_{j}  - p_{i} \chi_{j} = ih/2\pi \delta_{ij}  $.

 This makes consistent the non commutative and symplectic structures. The SU(2) version of the gravity gauge theory expressed as a holonomy in loop space is possible to construct in the commutative geometry limit. The field $\chi $ is chosen to have the exact and closed form conditions that  fail at the vortex like topological defects. The length scale for the non commutative structure is  chosen  small enough for the spacetime to be normal almost everywhere to do the usual physics. Are topologically non trivial states consistent with the standard general relativity using commutative geometry and is the use of non commutative structures the only way , or one of the ways, for them to occur is a unresolved question. The consistency of the theory of Yang-Mills type of all fields with non commutative geometry is being established from its inception.There are likely to be connections between loop space quantum gravity and quantum Riemannian geometry with non commutative geometry.

 The characterisation of the topological states arising in the cosmological models is done  with conditions on the vector field introduced for non commutative geometry. In differential geometry usually the Stokes law would read $ \int_{M} d\omega  = \int_{\partial M} \omega $, and $d^{2}\chi = 0$ every where except at the topological defects. The $ \chi $ field as a vortex like field will have $d^{2} \chi \neq 0$ ; $ \omega = d\chi$. This fails at the topological defects. The non commutative structure suggests that these defects are vortex like.  The vortex has a 'length'  scale given by $l$, and the $\theta $ is a skew symmetric form or matrix, so diagonal terms are zero. If the other terms also equal zero then those fields or coordinates commute. The planar vortex for example is described as two non commuting variables, with a $ \theta_{12} $ form, such that the exterior derivative $ d^{2} $  is non zero with an exclusion at the origin, such as $(\chi_{1} d\chi_{2} -\chi_{2} d\chi_{1})*((\chi_{1})^{2} +(\chi_{2})^{2})^{-1}$. In usual vector calculus the  curl of the  gradient of a scalar field and the  divergence of the curl of a vector field are zero. Or as expressed in terms of differential forms the $d^{2} $ is zero. The de Rham integration on non simply connected spaces and those with homotopy and cohomology ,  gives  the condition for the failure of these differential and integral conditions. This indicates presence of vorticity and topologically non trivial states or'defects'.

Consider a simple example of a holonomy loop of the U(1) group  around the vortex like topological defect : $ Trace P exp(-\oint A dx)$ . That is the 'pinched and popped' defect in space time which is the dark matter candidate. The length scale of $ 10^{-27} $ meters for the loop  is associated with the likely mass of $ 10^{-15} $kg ; and the gauge field $ A $ is the massless multiplet of GUT gauge fields at $ 10^{15} $ Gev.  The homotopy group $\pi_{1} (S^{1}) =Z $ and $\pi_{1}(SU(N)) = Z $. Higher homotopy groups for higher dimension spheres could be considered. The  fundamental group gives the set of integers and a winding number interpretation. A suitable Hopf index can be obtained from the $\oint \omega \wedge d\omega $ integral. This is a topological invariant on the 3 spheres in space time. The confinement of $ 10^{-15} $ kilograms to a region of size $ 10^{-27} $ meters in a topologically 'excluded' part of the spacetime manifold, does not create collapse to a black hole as the Schwarzschild radius for that would be around $ 10^{-40} $ meters. This mass energy can be released into the surrounding normal commutative spacetime  when the defect decays quantum theoretically into graviton and  ultra relativistic (GUT scale) effectively massless gauge bosons.  In the holonomy representation these  are represented by a U(1) gauge photon, as a non commutative vortex state is destroyed for which the fundamental group was Z, the group of integers.

The abundance of dark matter expected in the universe implies that these 'defects' occur abundantly, and that the tiny length scales of the loops around them means that for the most part smooth geometry prevails and general coordinate transformations required for Einstein's Theory of general relativity, remain valid everywhere except in these topologically excluded regions in which most mass/energy resides. The nature of dark energy is less certain and could be the cosmological term in Einstein's theory and hence at length scales much bigger than the Planck scale or the noncommutative vortex scales. Another major question is the role of the scalar field  that is used in inflation models and the Higgs scalar field used in particle physics  in these  topological solutions. Will it be neccessary to have additional fields to describe dark matter and energy , or are  they  already contained in the theory, with solutions including the non commutative structure. The reheating phase after the rapid cooling phase in inflation models  may not destroy the topological states , although it is supposed to smooth out other anisotropies and inhomogenities.

Vortex formation is associated with angular momentum and the spin network of the quantum space time foam indeed gives the quantised geometric operators of area nad volume in Quantum Riemannian geometry as  for example, area proportional to $(l_{p})^{2}  (j(j+1))^{1/2} $. From the Planck scale $l_{p} =10^{-35} $ meters to the non commutative scale of $ l= 10^{-27} $ meters, the continuum limit is reached ; and derivation ,commutation operations can be defined. This is taken at the upper limit in energy  of the GUT scale with  the $ 0.033 $  asymptotic converging coupling constant . This is  because operationally, the probes for finer grain topology of the space time can be only the GUT scale ; as our measurement methods depend on particle states with respect to locally Lorentzian spacetime. If astronomical observations also give the gamma energies corresponding to a decaying topological defect of this scale, then there is  a consistency. The defects of walls, strings, monopoles and texture type have been considered in other work and found to have limitations in explaining the dark matter and decay into high energy gamma rays.

Conclusion

The correspondence with other  theories could be made for (a) Vortex and  topological defect formation in field theory and condensed matter theories, (b) Quantum Riemann geometry and loop  space quantum Yang Mills  field theory  and quantum gravity theories, (c) Non commutative geometry theories for field theory and condensed matter theory,(d) Inflationary and model cosmologies that give the anisotropy and matter creation observed  in the universe.

Acknowledgements

I thank The Director, Dr Balasubramaniam  and  the Faculty and staff of the Institute of Mathematical Sciences , Chennai, India  for supporting  my academic visits to the Institute and providing the facilities ; that made  four  papers possible, listed in arxiv.org quant-ph.  The interest of  Professors Hari Dass, Simon, Govindarajan, Baskaran, Sharatchandra  and Date in my work is particularly acknowledged. I also appreciated the lectures in the Institute, and the discussions with Drs Menon, Shankar, Ganguly, , Balakrishnan, Divakaran, Rajasekaran ; and other members, visitors and students at the Institute.

References

(1)  John Peacock, Cosmological physics, Cambridge University Press (1998)

(2)  Giuseppe Morandi, The role of Topology in classical and quantum physics, Springer Verlag (1992)

(3)  Mikio  Nakahara,  Geometry, Topology and Physics, Institute of Physics publication (1990)

(4)  Theodor Fraenkel , Geometry of Physics ,Cambridge University press (1995)

(5)  David Thouless, Topological quantum numbers in non relativistic physics, World Scientific press(1998)

(6)  Shahn Majid , Quantum groups and non commutative geometry,arxiv, hep-th/0006167

(7)  Ivanhoe Pestov, Geometry of Manifolds and dark matter Ed.S.Dimiev and S.K.Sekigawa, WSP 2001 gr-qc/0212037

(8)  Ajay Patwardhan, Particle and Field symmetries and non commutative geometry,arxiv.org quant-ph 0305150

(90  Ajay Patwardhan, Quantisation in general spaces arxiv quant-ph 0211039

(10) Orlando Alvarez, Lectures on Quantum Mechanics and the Index Theorem

(11) Lee,Brekke,S.J.Collins,T.D.Imbo, Non Abelian vortices on surfaces and their statistics, Nuclear Physics B 500 (FS) (1997) 465-485

(12) Alain Connes A short survey of noncommutative geometry, Journal of Mathematical physics vol41,\#6, June 2000,3832-3866.

(13) Daniel Kastler , Non commutative geometry and fundamental physical interactions, Journal of Mathematical Physics ,volume 41,\#6 June 2000,3867-3891.

(16) A.Pinzul and A.Stern , A new class of two dimensional non commutative spaces, arxiv hep-th/0112220.

(17) H.O.Girotti Non commutative field theories arxiv , hep-th/0301237.

(18) Joseph Varilly, An introduction tomnon commutative geometry, arxiv physics/9709045.

(19) H.Grosse , C. Klimcik , P.Presnajder, Topologically Non trivial Field configurations in non commutative geometry, arxiv hep-th/9510083.
 
(20) C.Sochichiu ,Guage invariance and Non commutativity arxiv hep-th/0202014

(21) J.A.Wheeler , Geometrodynamics ,Academic press (1962)

(22) Abhay Ashtekar,Quantum Mechanics of Geometry,arxiv gr-qc/9901023

(23) Abhay Ashtekar, Quantum Geometry and Gravity,arxiv gr-qc/0112038

(23) Abhay Ashtekar, Quantum Theory of Geometry: Non commutativity of Riemannian structures. ,arxiv gr-qc/9806041

(24) M. Reuter , Non commutative geometry on quantum phase space,arxiv hep-th/9510011

(25) J.Frohlich, O.Grandjean and A,Recknagel , Supersymmetric quantum theory and Non commutative geometry, arxiv math-ph/9807006

(26) N.Seiberg and E.Witten, String theory and Non commutative geometry,arxiv hep-th/9908142

(27) Vilenkin and Shellard, Cosmic strings and topological defects, Cambridge university press(1995)

(28) Robert Brandenberger, Topological defects and cosmology,arxiv hep-ph/9806473; Pramana 51,1998,191-204

(29) Shinji Tsujikawa , Introductory review of Cosmic Inflation,arxiv hep-ph/0304257

(30) Houri Ziaeepour, A Decaying ultra heavy dark matter :Review of recent progress, arxiv astro-ph/0005299

(31) R.Durrer, M.Kunz, A.Melchioni ,Cosmic structure formation with topological defects,arxiv astro-ph/0110348, Physics reports v364,(2002),p1-81.

(32) S.Alexander, J Magueijo, Non commutative geometry as realisation of varying speed of light cosmology, arxiv hep-th/0104093

(33) Alejandro Gangui ,astro-ph /0110285

(34) arxiv astro-ph/0005299 and 9811011

(35) J.Madore, Introduction to Non commutative differential geometry  and applications to physical sciences, Cambridge Univerity press (1998)

(36) Ajay Patwardhan ,Classical and quantum phase space , separability and entanglement, arxiv quant-ph 0211041

\end{document}